\documentclass[aps,pre,onecolumn,superscriptaddress,showpacs]{revtex4-1}
\usepackage{graphicx}
\usepackage{dcolumn}
\usepackage{bm}

\usepackage{natbib}
\usepackage[
text={7in,9.5in},centering,
]{geometry}

\usepackage{float}
\usepackage{epsfig}
\usepackage{amsmath}
\usepackage{amssymb}
\usepackage{textcomp}
\usepackage{layouts}

\begin{document}

\title{Feynman Propagator for Interacting Electrons in the Quantum Fokker Theory}

\author{Natalia Gorobey}
\affiliation{Peter the Great Saint Petersburg Polytechnic University, Polytekhnicheskaya
29, 195251, St. Petersburg, Russia}

\author{Alexander Lukyanenko}\email{alex.lukyan@mail.ru}
\affiliation{Peter the Great Saint Petersburg Polytechnic University, Polytekhnicheskaya
29, 195251, St. Petersburg, Russia}

\author{A. V. Goltsev}
\affiliation{Ioffe Physical- Technical Institute, Polytekhnicheskaya
26, 195251, St. Petersburg, Russia}

\begin{abstract}
A modification of the Fokker action is proposed, which allows one to formulate the covariant quantum theory of the charge system, in which the proper time of each particle serves as the evolution parameter and the particles themselves receive an additional quantum number - spin. The modification consists in adding to the Fokker action its variation generated by the infinitesimal shifts of the proper time parameters. As a result, the proper time parameters become observable at the quantum level.

\end{abstract}


\maketitle

\section{Introduction}

Wheeler and Feynman \cite{1,2} proposed the formulation of classical electrodynamics without using an electromagnetic field, based on the Fokker theory \cite{3}. However, a quantum version of this theory has not been proposed. The reason is that the Fokker action does not have the usual Lagrangian form, which means that the usual rules of canonical quantization cannot be applied to it. For the simplest system of two particles, this action (in a parameterized form) has the form ($c=1$):
\begin{equation}
 I_F=\frac{1}{2} \int_0^1 d\tau_1(\frac{\dot{x}_1^2}{N_1}+ m_{1}^2 N_1)+\frac{1}{2} \int_0^1 d\tau_1(\frac{{x'_2}^2}{N_2}+m_2^2 N_2)+I_{int},
 \label{1}
\end{equation}

\begin{eqnarray}
I_{int}=\int_0^1 e_1 d\tau_1 \int_0^1 e_2 d\tau_2 \delta(s^{2}_{12}) \dot{x}_1 x'_{2}
\notag \\
  +\frac{1}{2} \int_{0}^1 e_{1} d\tau_1 \int_0^1 e_{\tilde1} d \tau_{\tilde1} \delta(s^2_{1 \tilde1}) \dot{x}_1 \dot{x}_{\tilde1}
  \notag \\
 +\frac{1}{2} \int_0^1 e_2 d\tau_2 \int_0^1 e_{\tilde2} d \tau_{\tilde2} \delta(s^2_{2 \tilde1}) {x}'_2 x'_{\tilde2}.
 \label{2}
\end{eqnarray}

The dot denotes the derivative with respect to the parameter $\tau_1$, and the prime denotes the derivative with respect to the parameter $\tau_2$. Repeating particle indices are indicated by a tilde. The following abbreviations for scalars in Minkowski space are used,
\begin{equation}
 \dot{x}_1 x'_{2} = \dot{x}_{1 \mu} {x'}_{2}^\mu, \dot{x}_1 \dot{x}_{\tilde1} = \dot{x}_{1 \mu} \dot{x}_1^{\mu}, x'_2 {x'}_{\tilde2}=x'_{2 \mu} {x'}_{\tilde2}^\mu,
  \label{3}
\end{equation}
as well as for invariants,
\begin{equation}
 s_{12}^2=(x_1-x_2)^2, s^2_{1 \tilde1}=(x_1-x_{\tilde1})^2, s^2_{2  \tilde2}=(x_2-x_{\tilde2})^2.
  \label{4}
\end{equation}
We included the self-interaction of particles, bearing in mind their possible reversal of Minkowski time $x^0$ in quantum theory. We also consider the charges of particles $e_1,e_2$ as a function of their proper time,
\begin{equation}
 s_1=\int_0^{\tau_1}N_{1}(\tau_1) d\tau_1,  s_2=\int_0^{\tau_2}N_{2}(\tau_{2})d\tau_2,
  \label{5}
\end{equation}
which will give us the opportunity to turn on and off the interaction (\ref{2}). Proper time (\ref{5}) is an invariant of reparametrizations:
\begin{equation}
\delta N_1=-\dot{\epsilon}_1, \delta N_2=-{\epsilon}'.
 \label{6}
\end{equation}
Thus, particle dynamics is described by two independent proper time parameters (\ref{5}), and interaction (\ref{2}) is nonlocal in these parameters. This partially motivated the introduction of a new formulation of quantum mechanics based on the functional integral in the configuration space of the system \cite{4}, which was used by Feynman to find the energy of the ground state of a polaron with time-nonlocal interaction \cite{5}. A more consistent definition of the integral, which is based on the operator formulation of quantum mechanics, leads to a functional integral in the phase space \cite{6}.
The presence of symmetries in theory necessitates a covariant formulation of the functional integral \cite{7}-\cite{11}, in which the reparameterization invariance of the proper time leads to additional integration over this parameter \cite{12}. Such integrals over proper time were introduced by Fock \cite{13} and Schwinger \cite{14} in the representation of Green's functions for the Dirac and Klein-Gordon operators - Feynman propagators. Thus, the functional integral in the covariant formulation of quantum theory gives the Feynman propagator, which in relativistic quantum particle mechanics does not have a direct dynamic meaning, but is a skeletal element of the diagram technique of the perturbation theory \cite{15}. In \cite{16}, the Fokker covariant quantum theory was formulated for spinless particles and its modification, which allows one to remove the integration over the particle´s proper time. The functional integral is introduced in the generalized phase space, in which the world lines of particles as a whole serve as generalized dynamic variables \cite{17,18}.
Modification of the theory is carried out at the classical level and consists in fixing the classical dynamics of the Minkowski space parameter $x^0$ for each particle in quantum mechanics. In this paper, a similar modification of covariant quantum theory is proposed for particles with spin, which we will call electrons in the following. It turns out that for this it is enough to fix the classical dynamics of the proper time (\ref{4}) of each particle, and interpret the resulting functional integral as it was done by Dirac when writing the wave equation for the electron.

\section{Modified Fokker action}
In \cite{16}, the modification of action (\ref{1}) consisted in adding to it variations in the coordinates $x^0$ of particles in Minkowski space generated by the infinitesimal shift of the proper time of each particle of the form (\ref{6}). As a result, this allows one to remove FS integration over proper time in the covariant Feynman propagator for scalar particles. For particles with spin (electrons), it is enough to fix the classical dynamics of their proper time (\ref{5}). This is achieved by adding to the initial action (\ref{1}) the variation generated by the infinitesimal reparametrizations (\ref{6}) (action (\ref{1}) is invariant with simultaneous compensating variation of all dynamic variables). Thus, the modified Fokker action for electrons has the form:
\begin{eqnarray}
 {\tilde{I}}_F =\frac{1}{2} \int_0^1 d\tau_1 [\frac{{\dot{x}}_1^2}{N_1} (1+\frac{{\dot{\epsilon}}_1} {N_1})+m_1^2 N_1]+\frac{1}{2} \int_0^1 d\tau_1[\frac{{x'}_2^2}{N_2}(1+\frac{{\epsilon}'_2}{N_2})+m_2^2 N_2]
\notag \\
+I_{int} +\int_0^1 d\tau_1 {\epsilon}_1 {\frac{de_1}{ds_1}} W_1+\int_0^1 d\tau_2 {\epsilon}_2 {\frac{de_2}{ds_2}} W_2,
 \label{7}
\end{eqnarray}
where
\begin{equation}
 W_1 = \int_0^1 e_{\tilde1} d\tau_{\tilde1} \delta(s^2_{1 \tilde1}) \dot{x}_1 \dot{x}_{\tilde1}+\int_0^1 e_2 d\tau_2 \delta(s^2_{12}) \dot{x}_1 {x'}_2,
  \label{8}
\end{equation}
\begin{equation}
 W_{2} = \int_0^1 e_{\tilde2} d\tau_{\tilde2} \delta(s^2_{1 \tilde2}){x'}_2 {x'}_{\tilde2} + \int_0^1 e_1 d\tau_{1} \delta (s^2_{12}) {\dot{x}}_1 {x'}_2.
  \label{9}
\end{equation}
We turn to the generalized canonical form of action (\ref{7}) in the same way as it was done in \cite{16}. We find generalized canonical momenta,
\begin{equation}
p_{1 \mu(s_1)}={\dot{x}}_{1\mu}(1+{\dot{\epsilon}}_1)+R_{1\mu},
 \label{10}
\end{equation}
\begin{equation}
 p_{2 \nu(s_2)}={x'}_{2\nu}(1+{{\epsilon}'}_2)+R_{2\nu},
  \label{11}
\end{equation}
where
\begin{eqnarray}
 R_{1\mu}=(e_1+2 {\epsilon}_1 \frac{de_1}{ds_1}) \int_0^{S_1} e_1 ds_1 \delta(s^2_{1 \tilde1}) \dot{x}_{1\mu}
 \notag \\
 +(e_1 +{\epsilon}_1 \frac{de_1}{ds_1}) \int_0^{S_2} e_2 ds_2 \delta(s^2_{12}){x'}_{2\mu}+ e_1 \int_0^{S_2}{\epsilon}_2 \frac{de_2}{ds_2} ds_2 \delta(s^2_{12}){x'}_{2\mu},
  \label{12}
\end{eqnarray}
\begin{eqnarray}
 R_{2\nu}=(e_2+2{\epsilon}_2 \frac{de_2}{ds_2}) \int_0^{S_2} e_{\tilde2}ds_{\tilde2} \delta(s^2_{2 \tilde2}){x'}_{\tilde2 \nu}
\notag \\
 +(e_2+\epsilon_{2} \frac{de_2}{ds_2}) \int_0^{S_1} e_1 ds_1 \delta(s^2_{12}) {\dot{x}}_{1\nu}+e_2 \int_0^{S_1} \epsilon_1 \frac{de_1}{ds_1} ds_{1} \delta(s^2_{12}){\dot{x}}_{1 \nu}.
  \label{13}
\end{eqnarray}
\begin{equation}
 P_{\epsilon_1}=\frac{1}{2}{\dot{x}}_1^2,
  \label{14}
\end{equation}
\begin{equation}
 P_{\epsilon_2}=\frac{1}{2}{x'}_2^2,
  \label{15}
\end{equation}
Hereinafter, we pass to the parameterization of the proper time of particles. We find the generalized Hamiltonian by means of the generalized Legendre transformation,
\begin{eqnarray}
 {\tilde{H}}_F = \int_0^{S_1} ds_1 p_1 {\dot{x}}_1 + \int_0^{S_2} ds_2 p_2 {x'}_2 - {\tilde{I}}_F
\notag \\
 =\int_0^{S_1}{ds_1}{P_{\epsilon_1}}(1+2{\dot{\epsilon}}_1)+\int_0^{S_2}{ds_2}{P_{\epsilon_2}}(1+2{\epsilon'}_2)-\frac{1}{2} m_1^2 {S_1}-\frac{1}{2}m_2^2 {S_2}
\notag \\
 +I_{int} +\int_0^1 d\tau_1 {\epsilon}_1 \frac{de_1}{ds_1} W_1+\int_0^1 d\tau_2 {\epsilon}_2 \frac{de_2}{ds_2} W_2,
  \label{16}
\end{eqnarray}
where the generalized velocities should be expressed in terms of the generalized momenta as a result of solving equations (\ref{10}) - (\ref{13}). In the framework of the perturbation theory (up to the first order), we find from (\ref{10}) and (\ref{11}):
\begin{equation}
 \dot{x}_{1\mu}(1+{\dot{\epsilon}}_1)= p_{1\mu}-R_{1\mu}[{p_1},{p_2}],
  \label{17}
\end{equation}
\begin{equation}
 {x'}_{2\nu}(1+{\epsilon}'_2)=p_{2\nu}-R_{2\nu}[{p_1},{p_2}],
  \label{18}
\end{equation}
where now
\begin{eqnarray}
 R_{1\mu} [p_1,p_2]=(e_1+2{\epsilon}_{1} \frac{de_1}{ds_1}) \int_0^{S_1} e_1 \frac{ds_1}{1+\dot{\epsilon}_1} \delta(s^2_{1\tilde1}) p_{1\mu}
\notag \\
 +(e_1+{{\epsilon}_1} \frac{de_1}{ds_1}) \int_0^{S_2} e_2 \frac{ds_2}{1+{\epsilon}'_2} \delta(s^2_{12}) p_{2\mu} + e_1 \int_0^{S_2} {\epsilon}_2 \frac{de_2}{ds_2} \frac{ds_2}{1+{\epsilon}'_2} \delta(s^2_{12}) p_{2\mu}+...,
  \label{19}
\end{eqnarray}
\begin{eqnarray}
 R_{2\nu} [p_1,p_2]=(e_2+2{\epsilon}_2 \frac{de_2}{ds_2}) \int_0^{S_2} e_{\tilde2} \frac{ds_{\tilde2}}{1+{\epsilon}'_{\tilde2}} \delta(s^2_{2 \tilde2}) p_{\tilde2\nu}
\notag \\
 +(e_2+\epsilon_2 \frac{de_2}{ds_2}) \int_0^{S_1} e_1 \frac{ds_1}{1+\dot{\epsilon}_1} \delta(s^2_{12}) p_{1\nu}
 + e_2 \int_0^{S_1} \epsilon_1 \frac{de_1}{ds_1} \frac{ds_1}{1+\dot{\epsilon}_1} \delta(s^2_{12}) p_{1\nu}+....
  \label{20}
\end{eqnarray}
From here, taking into account (\ref{14}) and (\ref{15}), we find:
\begin{equation}
 \dot{\epsilon}_1=-1+\frac{\sqrt{(p_1-R_1)^2}}{\sqrt{2P_{{\epsilon}_1}}},
  \label{21}
\end{equation}
\begin{equation}
 {{\epsilon}'}_2=-1+\frac{\sqrt{(p_2-R_2)^2}}{\sqrt{2P_{{\epsilon}_2}}}.
  \label{22}
\end{equation}
We note that the velocities $\dot{\epsilon}_1$, ${{\epsilon}'}_2$ are not yet determined by equations (\ref{21}), (\ref{22}), since they are still contained in $R_1,R_2$. Nevertheless, we write the generalized Hamiltonian in the form
\begin{eqnarray}
 \tilde{H}_F=\int_0^{S_1}ds_1[-P_{\epsilon_1}+\sqrt{2P_{\epsilon_1}}\sqrt{(p_1-R_1)^2}]+\int_0^{S_2}ds_2[-P_{\epsilon_2}+\sqrt{2P_{\epsilon_2}}\sqrt{(p_2-R_2)^2}]
\notag \\
 +I_{int} +\int_0^1 d\tau_1 \epsilon_1 \frac{de_1}{ds_1} W_1 + \int_0^1 d\tau_2 \epsilon_2 \frac{de_2}{ds_2} W_2-\frac{1}{2} m_1^2{S_1}-\frac{1}{2} m_2^2{S_2},
  \label{23}
\end{eqnarray}
where now the interaction terms are expressed through momenta as a result of substitution there $\dot{x}_{1\mu}, x'_{2\nu}$ according to (\ref{17}) and (\ref{18}). Since the generalized Hamiltonian (24) contains velocities $\dot{\epsilon}_1,\epsilon'_2$ that are not yet defined, we add additional conditions
\begin{equation}
 \varphi_1(\dot{\epsilon}_1,\epsilon_1,\epsilon_2)\equiv2 P_{\epsilon_1}(1+\dot{\epsilon}_1)^2-(p_1-R_1)^2,
  \label{24}
\end{equation}
\begin{equation}
 \varphi_2(\epsilon'_2,\epsilon_1,\epsilon_2)\equiv2P_{\epsilon_2}(1+\epsilon'_2)^2-(p_2-R_2)^2,
  \label{25}
\end{equation}
to the generalized canonical action with the corresponding Lagrangian multipliers $\lambda_1,\lambda_2$ by $\dot{\epsilon}_1\equiv\mu_1,\epsilon'_2\equiv\mu_2$. These conditions are quadratic with respect to the infinitesimal shifts. Since we need linearity of action along these shifts $\epsilon_1,\epsilon_2$, we add two more conditions
\begin{equation}
 \varphi_3=\epsilon_1-\eta_1=0,
  \label{26}
\end{equation}
\begin{equation}
 \varphi_4=\epsilon_2-\eta_2=0,
  \label{27}
\end{equation}
with the corresponding Lagrangian factors $\lambda_3,\lambda_4$. Thus, the modified Fokker action for electrons in a generalized canonical form has the form:
\begin{eqnarray}
 \tilde{I}_F=\int_0^{S_1}ds_1p_1\dot{x}_1+\int_0^{S_2}ds_2p_2x'_2-\tilde{H}_F
\notag \\ +\lambda_1\varphi_1(\mu_1,\eta_1,\eta_2)+\lambda_2\varphi_2(\mu_2,\eta_1,\eta_2)+\lambda_3\varphi_3(\epsilon_1,\eta_1)+\lambda_4\varphi_4(\epsilon_2,\eta_2).
 \label{28}
\end{eqnarray}

\section{Feynman propagator}
We immediately write down the formal representation for the Feynman propagator in the form of a functional integral in a generalized phase space of the modified system (we omit all constant factors in the integral measure):
\begin{eqnarray}
 \Phi=\int_0^{\infty} dS_1 \int_0^{\infty} dS_2 \int \prod\limits_{s_1\mu} dp_1dx_1 \prod\limits_{s_2\nu} dp_2dx_2\prod\limits_{s_1} dP_{\epsilon_1}d\epsilon_1 \prod\limits_{s_2} dP_{\epsilon_2}d{\epsilon_2}
\notag \\
 \times\prod\limits_{s_1} \frac{d\lambda_1 d\mu_1}{[P_{\epsilon_1}(1+\mu_1)]^{-1}} \prod\limits_{s_2}\frac{d\lambda_2d\mu_2} {[P_{\epsilon_2}(1+\mu_2)]^{-1}}\prod\limits_{s_1}d\lambda_3d\eta_1\prod\limits_{s_2}d\lambda_4 d\eta_2 \exp({\frac{i}{\hbar}\tilde{I}_F}).
  \label{29}
\end{eqnarray}
Here we also took into account additional conditions (\ref{24})-(\ref{27}), and the measure of integration over the variables $\lambda_1,\mu_1$ and $\lambda_2,\mu_2$ indicates that we limited ourselves to extracting the square roots (\ref{21}) and (\ref{22}), and velocities $\dot{\epsilon}_1,\epsilon'$ are still need to be determined. The  conditions are needed only starting from the first order of the perturbation theory. Now we have two tasks: to give the formal expression (\ref{29}) the physical meaning of the evolution operator with respect to the parameters of the proper time of the particles, as well as the matrix sense in the space of electron spin states. This will be accompanied by a consistent reduction in integrations in (\ref{29}).
Let's start with the first task. Integration over variables $\epsilon_1,\epsilon_2$ gives a product of $\delta$-functions
\begin{equation}
 \delta(\dot{P}_{\epsilon_1}+\dot{\epsilon}_1 W_1+\lambda_3),
  \label{30}
\end{equation}
\begin{equation}
 \delta(\dot{P}_{\epsilon_2}+\dot{\epsilon}_2 W_2+\lambda_4),
  \label{31}
\end{equation}
in which differential equations of changes observables $P_{\epsilon_1},P_{\epsilon_2}$ during the evolution of the system are concluded. It is clear why we need to turn interaction on and off: otherwise, these observables remain constant and we do not have any measure of proper time. In the absence of interaction, we put
\begin{equation}
 P_{\epsilon_1}=\frac{m_1^2}{2}, P_{\epsilon_2}=\frac{m_2^2}{2}.
  \label{32}
\end{equation}
The moments of switching on and off can be compared with the corresponding values of the coordinates $x^0$ of the Minkowski space, so that these coordinates will play the role of physical time. Integration over observables $P_{{\epsilon}_1},P_{{\epsilon}_2}$ is equivalent to solving evolution equations for them. In this case, there remain two $\delta$-functions containing the first integrals of the equations of evolution, depending on the interval of the proper time between switching on and off the interaction. It is these $\delta$-functions that remove FS integration over the parameters of proper time and give dynamic meaning to the Feynman propagator (\ref{29}).

Now let's get down to the second task. It is similar to that which Dirac decided in deriving his wave equation for an electron. We will use his solution here, defining the square roots in the generalized Hamiltonian (\ref{23}) as follows:
\begin{equation}
 \sqrt{(p_1-R_1)^2}=\hat{\gamma}_1^\mu (p_{1\mu}-R_{1\mu}), \sqrt{(p_2-R_2)^2}=\hat{\gamma}_2^\nu(p_{2\nu}-R_{2\nu}),
  \label{33}
\end{equation}
\begin{equation}
 \hat{\gamma}_1^\mu={\hat{\gamma}^\mu}{\times}{\hat{E}_4},  \hat{\gamma}_2^\nu={\hat{E}_4}\times\hat{\gamma}^\nu\,
  \label{34}
\end{equation}
where $\hat{\gamma}^\mu$ are Dirac matrices \cite{15}, $\hat{E}_4$ - unit matrix.
Now the Feynman propagator (\ref{29}) must be defined as a matrix operator acting in the space of electron states, which are Dirac bi-spinors. First of all, we assume that the exponents are chronologically ordered by the parameters of their own time. Let's start with integration over the momenta. For fixed values of the electron proper time, we have a Gaussian integral over the momenta of the particles, which we immediately ``calculate'':
\begin{equation}
 \int\prod\limits_{\alpha} dp \exp[-\frac{i}{\hbar}(A_{\alpha\beta}p_\alpha p_\beta +p_\alpha \hat{\Gamma}_\alpha)]=\sqrt{\frac{2\pi\hbar}{idetA}} \exp[\frac{i}{4\hbar} A^{-1}_{\alpha\beta} \hat{\Gamma}_\alpha \hat{\Gamma}_\beta].
  \label{35}
\end{equation}
where $\hat{\Gamma}_\alpha$ are the matrices in the space of bi-spinors that linearly depend on the velocities $\dot{x}_1,x'_2$. We give this expression meaning in the framework of perturbation theory. In the zeroth order, when $e_{1,2}\rightarrow 0$, we have,
\begin{equation}
 p_\alpha \hat{\Gamma}_\alpha= {\hat{\gamma}}_1^\mu p_{1\mu} + \hat{\gamma}_2^\nu p_{2\nu}-p_{1\mu} {\dot{x}}_1^\mu-p_{2\nu}{x'}_2^\nu,
  \label{36}
\end{equation}
\begin{equation}
 A_{\alpha\beta} p_\alpha p_\beta =\lambda_1 p^2_1+\lambda_2 p_2^2,
  \label{37}
\end{equation}
where $\lambda_1,\lambda_2$ have first order by $e_1,e_2$. In this limit, integral (\ref{29}) reduces to the product of free Feynman propagators of electrons, each of which includes a functional integral in the phase space of the particle:
\begin{equation}
 \int\prod\limits_{s\mu} dp dx \exp[\frac{i}{\hbar}\int_0^S ds(p\dot{x}-\hat{H})],
  \label{38}
\end{equation}
with
\begin{equation}
 \hat{H}=m{\hat{\gamma}^\mu}p_\mu-m^2.
  \label{39}
\end{equation}
It is also obvious that it should be equal to the kernel of the evolution operator in the space of bi-spinors:
\begin{equation}
 \hat{T}=\exp(-\frac{i}{\hbar}\hat{H}S).
  \label{40}
\end{equation}
By setting the zero approximation of perturbation theory in this way, we determine the spin structure of the propagator (\ref{29}) by expanding the matrix exponent (\ref{35}) into a series of perturbation theory.
With such a definition of the evolution operator, the probabilistic interpretation of the electron wave function $\psi$ is usual: $\psi^*\psi$ is the probability density distribution in space for a fixed $x^0$.

\section{Conclusions}
We formulated the problem of scattering of two electrons (positrons, at the corresponding boundary values $x^0$) in the framework of the quantum Fokker theory, in which the degrees of freedom of the electromagnetic field are excluded. Remembering that the Fokker theory was adapted by Wheeler and Feynman for an alternative formulation of the classical theory of electromagnetic interaction, we naturally ask the question: how does this theory relate to ordinary quantum electrodynamics? In the formulation of Wheeler and Feynman, the concept of an absorber was important, which took into account the combined effect of all charges in the universe. This means that in the Fokker quantum theory one should also calculate (and average accordingly) the Feynman propagator for all charges in the universe. It can be expected that thereby the vacuum effects of the electromagnetic field predicted by quantum electrodynamics will be taken into account. At the same time, the proposed formalism allows one to take into account the effect of charges having closed world lines.
The dynamics of loop electrons is not modified, and FS integrals over their proper time remain. These contributions could correspond to the vacuum effects of the electron-positron field. To this, however, superposition rules for such contributions should be added.

\section{acknowledgements}
The authors thank V.A. Franke for useful discussions.


\begin{thebibliography}{99}
\bibitem{1} R. P. Feynman and J. A. Wheeler, Rev. Mod. Phys. \textbf{17}, 157 (1945).

\bibitem{2} R. P. Feynman and J. A. Wheeler, Rev. Mod. Phys. \textbf{21}, 425 (1949).

\bibitem{3} A. D. Fokker, Z. Phys. \textbf{58}, 386 (1929).

\bibitem{4} R. P. Feynman, The Development of the Space-Time View of QuantumElectrodynamics, Nobel Lecture, December 11, 1965. Preprint les PrixNobel en 1965. The Nobel Foundation. Stockholm, 1966.

\bibitem{5} R.P. Feynman and A. R. Hibbs, Quantum Mechanics and Path Integrals,McGraw-Hill, New-York, (1965).

\bibitem{6} L. D. Faddeev and A. A. Slavnov, Gauge fields: An introduction to quantum theory, Westview Press, 2d edition, 236 (1993).

\bibitem{7} E. S. Fradkin and G. A. Vilkovisky, Phys. Lett. 55B, 224 (1975).

\bibitem{8} I. A. Batalin and G. A. Vilkovisky, Phys. Lett. \textbf{69B}, 309 (1977).

\bibitem{9} M. Henneaux, Physics Reports \textbf{126}, 1 (1985).

\bibitem{10} T. Kugo and I. Ojima, Suppl. Progr. Theor. Phys. \textbf{66}, 1 (1979).

\bibitem{11} F. R. Ore and P. van Nieuwenhuisen, Nucl. Phys. B 204, 317 (1982).

\bibitem{12} Jan Govaerts, A note of the Fradkin-Vilkovisky theorem, CERN-TH 5010/88 (1988).

\bibitem{13} V.A. Fock, Izv. Acad. Nauk SSSR, 551 (1937).

\bibitem{14} J. Schwinger, On gauge invariance and vacuum polarization, Phys. Rev. \textbf{82}. N 5, 664-679 (1951).

\bibitem{15} James D.Bjorken, Sidney D.Drell, Relativistic quantum fields, Mc Graw-Hill Book Company (1978).

\bibitem{16} Natalia Gorobey, Alexander Lukyanenko, and A. V. Goltsev, arXiv:2002.03607v1 (2020).

\bibitem{17} X. Jaen, R. Jauregui, J. Llosa, and A. Molina, Phys. Rev. D \textbf{36}, 2385 (1987).

\bibitem{18} X. Jaen, R. Jauregui, J. Llosa, and A. Molina, J. Math. Phys. \textbf{30} (12), 2807-2814 (1989).
\end{thebibliography}
\end{document}